\documentclass[preprint,12pt]{elsarticle}

\usepackage{graphicx}
\usepackage{dcolumn}
\usepackage{bm}

\usepackage[utf8]{inputenc}
\usepackage[T1]{fontenc}
\usepackage{mathptmx}
\usepackage{etoolbox}

\makeatletter
\def\@email#1#2{%
 \endgroup
 \patchcmd{\titleblock@produce}
  {\frontmatter@RRAPformat}
  {\frontmatter@RRAPformat{\produce@RRAP{*#1\href{mailto:#2}{#2}}}\frontmatter@RRAPformat}
  {}{}
}%
\makeatother

\usepackage[margin=2cm]{geometry}

\usepackage{subfig}
\usepackage{hyperref}
\usepackage[dvipsnames]{xcolor}
\usepackage{amsmath}
\usepackage{amsfonts,amssymb}
\usepackage{tikz}
\usetikzlibrary{quantikz}
\usepackage{tabularx}
\usepackage{soul}  
\usepackage{pgfplots}
\pgfplotsset{compat=1.18}
\usepackage{mathbbol}
\usepackage{xifthen}
\usepackage[titletoc,title]{appendix}
\usepackage{multirow}

\newcommand{\yi}{\mathrm{i}}
\newcommand{\oO}{\mathcal{O}}
\newcommand{\ylb}{\left(}
\newcommand{\yrb}{\right)}
\newcommand{\yFi}{\mathcal{F}}
\DeclarePairedDelimiter\floor{\lfloor}{\rfloor}

\usepgfplotslibrary{fillbetween}

\journal{Journal of Computational Physics}
\begin{document}

\begin{frontmatter}

\title{Estimating QSVT angles for matrix inversion with large condition numbers}
\author[1]{I. Novikau}
\ead{novikau1@llnl.gov}
\author[1]{I. Joseph}
\affiliation[1]{organization={Lawrence Livermore National Laboratory},
            addressline={7000 East Ave}, 
            city={Livermore},
            postcode={94550}, 
            state={California},
            country={USA}}

\begin{abstract}
Quantum Singular Value Transformation (QSVT) is a state-of-the-art, near-optimal quantum algorithm that can be used for matrix inversion. 
The QSVT circuit is parameterized by a sequence of angles that must be pre-calculated classically, with the number of angles increasing as the matrix condition number grows.
Computing QSVT angles for ill-conditioned problems is a numerically challenging task. 
We propose a numerical technique for estimating QSVT angles for large condition numbers. 
This technique allows one to avoid expensive numerical computations of QSVT angles and to emulate QSVT circuits for solving ill-conditioned problems.
\end{abstract}

\begin{keyword}
Quantum computing \sep QSVT \sep Matrix inversion \sep Quantum Linear System Algorithm 
\end{keyword}
\end{frontmatter}

\section{Introduction}\label{sec:introduction}

\subsection{Motivation}
Quantum Singular Value Transformation (QSVT)~\cite{Gilyen19, Martyn21, Lin22} is a modern quantum algorithm used to perform Hamiltonian simulation, solve systems of linear equations, and, in general, compute polynomials of matrix singular values.
The shape of a polynomial computed by QSVT is determined by a sequence of angles $\phi$ that must be pre-calculated classically and parameterize the QSVT circuit.
In particular, the QSVT algorithm is a near-optimal method for matrix inversion and can, therefore, serve as a quantum linear solver.
In this context, the number of QSVT angles grows at least linearly with the matrix condition number $\rho$.
However, many systems of practical interest are stiff, with corresponding matrices characterized by high $\rho$.
In such cases, high-precision computation of $\phi$~\cite{Dong21, Ying22, Dong23} becomes a challenging numerical task already for $\rho$ around $10^3$.
Notably, the computational cost of these methods~\cite{Dong21, Ying22} grows quadratically with $\rho$.
The complexity was reduced to near-linear using the so-called Fast-Fixed-Point-Iteration (FFPI) technique recently proposed in Ref.~\cite{Ni24}, which significantly accelerates the high-precision computation of the QSVT angles.
In our work, we propose an alternative fast numerical technique to estimate QSVT angles for large condition numbers.
This approach allows QSVT to be applied to approximately solve ill-conditioned systems of linear equations, while significantly reducing the numerical resources required for computing QSVT angles.

The estimation algorithm proposed in this work was motivated by the following question: is it possible to reduce the optimization of a quantum circuit that depends on $N_{\rm pars}$ parameters to an optimization with respect to $N_{\rm meta} \ll N_{\rm pars}$ metaparameters?
Let us assume that we have a quantum circuit $C[p]$ computing an object function $f(x)$ depending on the variable $x$, and this circuit depends on $N_{\rm pars}$ parameters $p(x)$.
Then, the direct approach for constructing $C[p]$ for a particular $x$ is the minimization of the cost function $L(p) = |f - C[p]|$ to find the most optimal parameters $p$.
Instead, one can try to find $N_{\rm meta} \ll N_{\rm pars}$ metaparameters $m$ that capture the dependence of $p$ on $x$.
These metaparameters can thus be considered a compressed version of the original parameters $p$.
In this case, minimizing the new cost function $L(m) = |f - C[m]|$ should require far fewer numerical resources than minimizing $L(p)$.
We find such metaparameters for the QSVT circuit in the matrix-inversion problem and, by using them, estimate the QSVT angles.
However, it is important to emphasize that the proposed technique is based solely on observations of the specific behavior of the QSVT angles for matrix inversion, and its rigorous analytical justification remains an intriguing open question.

\subsection{Applications of Quantum Linear System Algorithms (QLSAs)}
Quantum Linear System Algorithms (QLSAs), such as QSVT, are used to solve a broad range of linear and nonlinear (NL) problems. 
In particular, QSVT was used to simulate stationary linear electromagnetic (EM) waves~\cite{Novikau23} and was proposed for modeling kinetic waves in an electron plasma~\cite{Novikau24-EVM}.  
It was also proposed in~\cite{Golden22} to use preconditioned QLSAs for the linear hydrological modeling, where it was shown that condition numbers of practical interest can reach $10^{6}$ or even larger values.
Similarly, it was described in~\cite{Henderson23, Henderson24} how QLSAs can be applied to modeling linear geological flows.
In general, QLSAs are also used to solve time-dependent differential equations~\cite{Berry24}.

On the other hand, NL problems linearized using the Carleman~\cite{Carleman32, Liu21, Krovi23} or Koopman--von Neumann~\cite{Koopman31, Joseph20, Dodin20} techniques can be simulated with QLSAs.
For instance, QLSA is a part of the Carleman-based algorithm discussed in~\cite{Liu21} for modeling dissipative dynamics with quadratic nonlinearity.
The same approach was proposed in~\cite{Akiba23} for solving first-order differential equations used in chemical kinetics.

Thus, QLSAs encompass a broad class of algorithms widely used for modeling both linear and NL dynamics, with QSVT being a well-known 
 and versatile QLSA.

\subsection{Notation}
Here, we summarize the main notation used throughout this work.
The scalar $\kappa_{\rm qsvt}$ is a parameter that defines an accurate approximation of the inverse function [Eq.~\eqref{eq:F}].
It is also an input parameter for the minimization algorithm~\cite{Dong21} described in Sec.~\ref{sec:minimization}, which computes high-precision QSVT angles $\phi$.
The scalar $\kappa_{\rm ref}$ is a specific value of $\kappa_{\rm qsvt}$ used to compute high-precision QSVT angles $\phi_{\rm ref}$, which serve as a reference dataset for the estimation technique presented in Sec.~\ref{sec:approach}.
The parameter $\kappa_0$ is the value of $\kappa_{\rm qsvt}$ used by the estimation algorithm to compute the estimated QSVT angles, denoted as $\bar{\phi}$.
The scalar $\rho_A$ is the condition number of the target matrix $A$ that needs to be inverted.

\subsection{Key results}
We propose a numerical technique for estimating QSVT angles for inverting matrices characterized by large condition numbers.
The computational time of this technique scales linearly with $\kappa_0$.
The algorithm is summarized in Table~\ref{table:algorithm}.
In particular, a block-encoded matrix $A$ can be inverted by the QSVT circuit with high precision using the QSVT angles $\phi$ computed by the minimization algorithm~\cite{Dong21} with $\kappa_{\rm qsvt} \geq ||A||^{-1}\rho_A$, where $||A||$ is the spectral norm of $A$. 
The maximum error $\varepsilon_{\rm qsvt}$ of these QSVT computations is determined by the chosen degree $N_c$ of the polynomial~\eqref{eq:series-Tk} approximating the inverse function~\eqref{eq:F}.
Similarly, the matrix $A$ can be inverted by the QSVT circuit using the estimated angles $\bar{\phi}$ computed for $\kappa_0 \geq ||A||^{-1}\rho_A$ with the estimation algorithm described in Sec.~\ref{sec:approach}.
Assuming the result of the inversion is normalized to one, we find that the maximum error $\varepsilon_{\rm appr}$ of the QSVT computations with the angles $\bar{\phi}$ gradually increases with $\kappa_0$ and is approximately $10^{-5}$ for $\kappa_0 \leq 10^6$ for matrices with $||A|| < 1$. 

This paper is organized as follows. 
In Sec.~\ref{sec:qsvt}, we review the QSVT algorithm for matrix inversion and the minimization approach for computing QSVT angles with high precision.
In Sec.~\ref{sec:approach}, we describe the numerical technique to estimate QSVT angles that can be used for inverting matrices with large condition numbers.
In Sec.~\ref{sec:tests}, we test the estimated angles on various examples including a linear boundary-value problem of stationary EM waves.
In Sec.~\ref{sec:conclusion}, we present the main conclusions.

\section{Overview of QSVT for matrix inversion}\label{sec:qsvt}

\subsection{Overview}\label{sec:qsvt-basics}

\begin{figure*}[!t]
\centering
\includegraphics[width=0.90\textwidth]{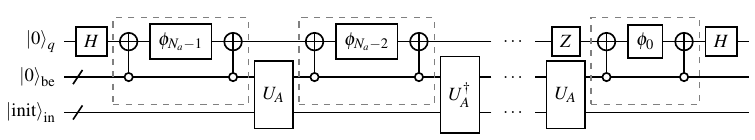}
\caption{
    \label{circ:qsvt} 
    The QSVT circuit computing $P[\phi](A)$ [Eq.~\eqref{eq:qsvt-odd}].
    Here, the gates schematically denoted as $\phi_j$ represent the rotations $R_z(2\phi_j)$.
}
\end{figure*}

Let us assume that we have an invertible $N\times N$ matrix $A$ whose singular value decomposition is
\begin{equation}\label{eq:svd}
    A = U_l S U_r^\dagger,
\end{equation}
where $U_l$ and $U_r$ are unitary matrices containing left and right singular vectors of $A$, correspondingly, and the matrix $S = {\rm diag}(s_0, \dots, s_{N-1})$ contains real strictly positive singular values $s_i$ of $A$.
QSVT~\cite{Gilyen19,Martyn21} builds a function $P(A)$ by transforming the singular values $s_i$:
\begin{equation}\label{eq:qsvt-definition}
    P(A) = U_l f(S) U_r^\dagger,
\end{equation}
where $f(S) = {\rm diag}(f(s_0), \dots, f(s_{N-1}))$, and it is assumed that $||A||\leq 1$, i.e. $s_{\rm max} \leq 1$.
(Further in the text, we use the small letter $s$ without any subindex to indicate either some singular value, or a set of singular values of $A$.)
In particular, the QSVT calculation of odd functions can be performed using the circuit shown in Fig.~\ref{circ:qsvt}. 
It consists of a sequence of $N_c+1$ parameterized rotations $\exp(\yi\phi_k Z_\Pi)$ (dashed gray boxes in Fig.~\ref{circ:qsvt}), alternating with $N_c$ calls to the oracle $U_A$, which block-encodes the target matrix $A$:
\begin{equation}\label{eq:qsvt-odd}
    P[\phi](A) = \bra{+}_q\bra{0}_{{\rm be}}\ylb 
        e^{\yi\phi_0 Z_\Pi} Z_q U_A e^{\yi\phi_1 Z_\Pi} \prod^{(N_c-1)/2}_{k=1} G_k
    \yrb \ket{+}_q\ket{0}_{{\rm be}},
\end{equation}
where
\begin{subequations}
\begin{eqnarray}
    &&G_k = U_A^\dagger e^{\yi\phi_{2k} Z_\Pi}  U_A e^{\yi\phi_{2k+1} Z_\Pi},\\
    &&\bra{0}_{\rm be} U_A \ket{0}_{\rm be} = A.\label{eq:UA}
\end{eqnarray}
\end{subequations}
The equation~\eqref{eq:qsvt-odd} and the corresponding circuit~\ref{circ:qsvt} compute the real polynomial $P[\phi](A)$~\cite{Lin22}.
Here, $Z_q$ is the Pauli $Z$ gate acting on the qubit $q$. 
The operator $Z_\Pi$ changes the sign of nonzero states of the ancillary register `be' used for encoding the matrix $A$.
The qubit $q$ serves as the target qubit for the controlled rotations $\exp(\yi\phi_k Z_\Pi)$ which perform either $\exp(\yi\phi_k)$ or $\exp(-\yi\phi_k)$ transformations depending on the state of the register `be'.

\begin{figure}[!t]
\centering
\includegraphics[width=0.49\textwidth]{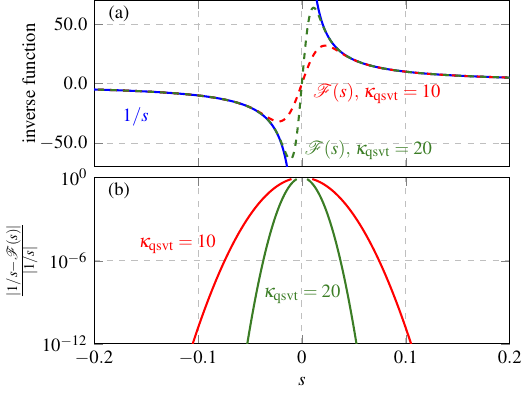}
\caption{
    \label{fig:inc-func} 
    (a) Plot showing the function $s^{-1}$ (blue solid line) and the function $\yFi(s)$ [Eq.~\eqref{eq:F}] with various $\kappa_{\rm qsvt}$ (dashed lines) constructed in the domain $s \in [-0.2, 0.2]$. 
    (b) The relative error in approximating $s^{-1}$ with the function $\yFi(s)$ with various $\kappa_{\rm qsvt}$.
}
\end{figure}

The notation $P[\phi]$ indicates that the QSVT circuit is parameterized by a set of angles $\phi$, precomputed classically.
These angles determine the shape of the function $f(s)$ in Eq.~\eqref{eq:qsvt-definition}.
We are interested only in the computation of the inverse function:
\begin{equation}\label{eq:f-inv}
    f(s) = s^{-1}.
\end{equation}
By approximating the function~\eqref{eq:f-inv}, QSVT inverts the original matrix~\eqref{eq:svd}.
Let us introduce the condition number $\rho_A$ of the matrix $A$:
\begin{equation}\label{eq:kappa}
    \rho_A = \frac{s_{\rm max}}{s_{\rm min}} = \frac{||A||}{s_{\rm min}}.
\end{equation}
Thus, $s_{\rm min} = ||A||\rho_A^{-1}$.
To avoid the singularity at $s = 0$ in Eq.~\eqref{eq:f-inv}, we approximate the inverse function~\eqref{eq:f-inv} by the odd function~\cite{Ying22},
\begin{equation}\label{eq:F}
    \yFi(s) = \frac{1 - e^{-(5s\kappa_{\rm qsvt})^2}}{s},
\end{equation}
which provides an $\epsilon$-approximation of the inverse function within the interval
\begin{equation}\label{eq:interv}
    [-1, -\kappa_{\rm qsvt}^{-1}] \cup [\kappa_{\rm qsvt}^{-1}, 1],
\end{equation} 
with $\epsilon \leq 10^{-12}$, as one can see in Fig.~\ref{fig:inc-func}.
To guarantee that the above interval includes the entire domain of the singular values of the matrix, $\kappa_{\rm qsvt}^{-1}$ must be less or equal than the minimal singular value of $A$ defined according to Eq.~\eqref{eq:kappa}:
\begin{equation}\label{eq:kappa-qsvt-norm}
    \kappa_{\rm qsvt} \geq ||A||^{-1}\rho_A = s_{\rm min}^{-1}.
\end{equation}
In this case, the function $\yFi(s)$ is $\epsilon$-close to the inverse function~\eqref{eq:f-inv} in the interval~\eqref{eq:interv} and is finite for $s \in [-\kappa_{\rm qsvt}^{-1}, \kappa_{\rm qsvt}^{-1}]$, as seen from Fig.~\ref{fig:inc-func}.
It is worth mentioning that one can use other approximations of the inverse function~\cite{Gilyen19, Dong21} where the error $\epsilon$ can be varied explicitly.

To construct the inverse matrix $A^{-1}$ by QSVT, one needs to precompute the QSVT angles $\phi$ setting $\yFi(s)$ as the target function as explained in Sec.~\ref{sec:minimization}.
Then, the optimized angles $\phi$ are used as the parameters of the QSVT circuit~\ref{circ:qsvt}.
The resulting circuit returns the quantum state
\begin{equation}\label{eq:qsvt-output}
    \ket{{\rm out}} = \ket{0}_{q,{\rm be}}\ket{\psi}_{\rm in} + \ket{\dots},
\end{equation}
where we are interested only in the state entangled with $\ket{0}_{q,{\rm be}}$,
\begin{equation}\label{eq:qsvt-output-res}
    \ket{\psi}_{\rm in} = \frac{\eta_{\rm qsvt} e^{\yi\phi_{\rm glob}}}{\kappa_{\rm qsvt}}A^{-1}\ket{{\rm init}}_{\rm in},
\end{equation}
where $\phi_{\rm glob}$ is an unknown global phase, $\eta_{\rm qsvt} < 1$ is a constant positive scalar that ensures smooth computation of QSVT angles by the minimization algorithm~\cite{Dong21, Dong21code}, and $\ket{{\rm init}}_{\rm in}$ is an initial state that needs to be multiplied by $A^{-1}$. 
In this work, we keep $\eta_{\rm qsvt} = 0.125$.

\subsection{Precise computation of QSVT angles}\label{sec:minimization}

Let us assume that $F(s)$ is the target function to be computed by QSVT.
In our case, we set $\eta_{\rm qsvt}\yFi(s)/\kappa_{\rm qsvt}$ as the target function.
To compute QSVT angles~\cite{Dong21, Dong21code}, first of all, the function $F(s)$ is approximated by a sequence of Chebyschev polynomials,
\begin{subequations}
\begin{eqnarray}
    &&\bar{F}(s) = \sum_{k=0}^{N_c} c_k T_k(s),\label{eq:series-Tk}\\
    &&\bar{F}(s) = F(s) + \varepsilon_{\rm qsvt},\label{eq:qsvt-err}
\end{eqnarray}
\end{subequations}
where $\varepsilon_{\rm qsvt}$ is the approximation error. 
The coefficients $c_k$ can be found using the Fourier series
\begin{equation}\label{eq:fourier-ck}
    c_k \approx \frac{2 - \delta_{k0}}{2N_q} (-1)^k \sum_{j = 0}^{2N_q - 1} F\ylb-\cos(j\pi/N_q)\yrb e^{\yi \frac{kj\pi}{N_q}},
\end{equation}
where $N_q$ should be not less than $N_c$, and the function $F(s)$ is assumed to be normalized such that its value is less or equal to one.
Since our target function~\eqref{eq:F} is odd, it can be approximated by an odd polynomial $\bar{F}$ with  
\begin{equation}\label{eq:Na-Nc}
    N_a = N_c + 1
\end{equation}
coefficients where $N_a$ is an even integer. 
On the other hand, for each $s$, the polynomial~\eqref{eq:series-Tk} can be constructed by using the following sequence of unitaries~\cite{Dong21}:
\begin{subequations}\label{eq:qsvt-comp}
\begin{eqnarray}
    &&U[\alpha](s) = e^{i\alpha_0 Z} \prod^{N_a-1}_{l=1} W(s) e^{\yi\alpha_l Z},\\
    &&W(s) = \begin{pmatrix}
            s  &   \yi \sqrt{1 - s^2}\\
            \yi\sqrt{1-s^2} & s
        \end{pmatrix},\\
    &&P[\alpha](s) = {\rm Re}\,U_{00}[\alpha](s),
\end{eqnarray}
\end{subequations}
which is another representation of the QSVT circuit for some singular value $s$.
Here, the angles $\alpha_j$ are shifted versions of $\phi_j$:
\begin{equation}\label{eq:alpha}
    \alpha_j = 
    \left\{ \begin{aligned}
        \phi_j - \pi/4&,\quad j = 0\text{ and } (N_a-1),\\
        \phi_j - \pi/2&,\quad j = 1, 2, \dots N_a-2.
    \end{aligned}\right.
\end{equation}

The algorithm for the computation of $\phi$ described in~\cite{Dong21} uses the fact that the QSVT angles $\phi$ satisfy the inversion symmetry, i.e. $\phi_j = \phi_{N_a - j - 1}$ for $j = 0,1,\dots,(N_a/2-1)$ (the same is true for $\alpha_j$).
Therefore, it is sufficient to compute only the first $N_a/2$ angles.
The algorithm calculates these angles by minimizing the following loss function
\begin{equation}\label{eq:loss-function}
    L(\alpha) = \sum_{k=0}^{N_a/2-1} \big|P[\alpha](x_k^{\rm Ch}) - \bar{F}(x_k^{\rm Ch})\big|^2,
\end{equation}
where $x^{\rm Ch}$ are Chebyschev roots, and the ansatz $\alpha_0 = \pi/4$ and $\alpha_j = 0$ for $j = 1,2,\dots (N_a/2-1)$ is taken.
If one sets the function~\eqref{eq:F} as the target function, then the minimization~\eqref{eq:loss-function} provides such angles $\alpha$ that
\begin{equation}\label{eq:P-F-comp}
    P[\alpha](s) = \frac{\eta_{\rm qsvt}}{\kappa_{\rm qsvt}}\yFi(s) + \varepsilon_{\rm qsvt}.
\end{equation}

\begin{figure*}[!t]
\centering
\subfloat[][QSVT angles]{\includegraphics[width=0.34\textwidth]{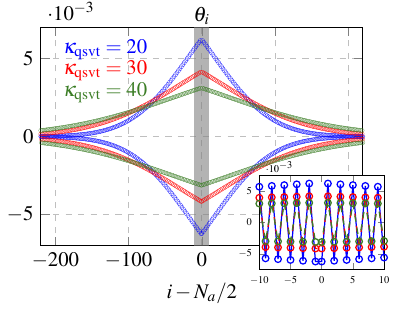}\label{fig:ref-angles}}
\subfloat[][$\max_j\theta_j$ versus $\kappa_{\rm qsvt}$]{\includegraphics[width=0.33\textwidth]{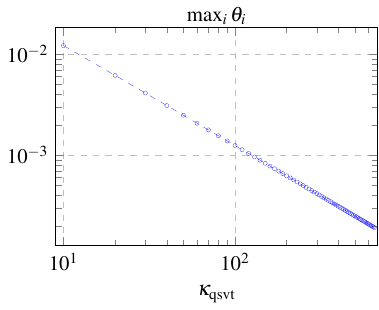}\label{fig:ref-ampl}}
\subfloat[][$N_a$ versus $\kappa_{\rm qsvt}$]{\includegraphics[width=0.31\textwidth]{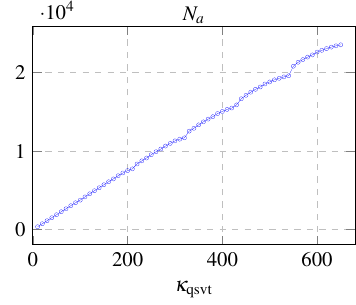}\label{fig:ref-Na}}
\caption{
    \label{fig:ref} 
    (a) Plot showing the shifted QSVT angles $\theta_i$ [Eq.~\eqref{eq:theta}] computed for various $\kappa_{\rm qsvt}$ using the minimization technique~\cite{Dong21}~\eqref{eq:loss-function} with $\varepsilon_{\rm qsvt} \approx 10^{-7}$. 
    Note that, for the chosen $\kappa_{\rm qsvt}$, only a subset of all angles is shown.
    The inner plot zooms in on $\theta_i$ within the shaded gray area. 
    (b) The dependence of $\max_i\theta_i$ on $\kappa_{\rm qsvt}$.
    (c) The dependence of the number of the QSVT angles on $\kappa_{\rm qsvt}$.
}
\end{figure*}

It is convenient to consider QSVT angles shifted by $\pi/2$:
\begin{equation}\label{eq:theta}
    \theta_j = \phi_j - \pi/2,\quad j = 0,1,\dots (N_a-1),
\end{equation}
which are similar to $\alpha_j$, except for the angles at $j = 0$ and $j = N_a-1$.
The angles $\theta$ computed by the minimization technique~\eqref{eq:loss-function} for various $\kappa_{\rm qsvt}$ are shown in Fig.~\ref{fig:ref-angles}.
One can see that the corresponding angles $\phi$ are symmetric with respect to $N_a/2-1/2$, their values oscillate around $\pi/2$, and their amplitudes tend toward $\pi/2$ for $|j - N_a/2| \gg 1$.
The maximum amplitude of $\theta$ is approximately inversely proportional to $\kappa_{\rm qsvt}$ (Fig.~\ref{fig:ref-ampl}).

The number $N_a$ depends on the target value of the parameter $\kappa_{\rm qsvt}$ and the error $\varepsilon_{\rm qsvt}$.
More precisely, the query complexity of QSVT scales~\cite{Gilyen19,Dong21} as $\oO(\kappa_{\rm qsvt}\log_2(\kappa_{\rm qsvt}\varepsilon_{\rm qsvt}^{-1}))$.
For a fixed $\varepsilon_{\rm qsvt}$, one can roughly assume that $N_a$ grows linearly with $\kappa_{\rm qsvt}$ (Fig.~\ref{fig:ref-Na}),
becoming significant for ill-conditioned problems.
Consequently, precise computation of $\phi$ for large condition numbers demands substantial numerical resources.
Yet, in practical simulations, it is often sufficient to estimate QSVT angles rather than compute them precisely. 
In the next section, we propose a simple numerical algorithm to accomplish this.

\section{QSVT angle estimation algorithm}\label{sec:approach}

Instead of directly computing $\phi$ for large condition numbers, we use the reference angles $\phi_{\rm ref}$ provided by the minimization procedure~\eqref{eq:loss-function} for a relatively low value $\kappa_{\rm ref}$ to estimate a set of metaparameters which describe the main features of the QSVT angles.
After that, using these metaparameters, we calculate the estimated angles $\bar{\alpha}$ (and the corresponding estimated angles $\bar{\phi}$ and $\bar{\theta}$) for the target value $\kappa_0$.
Our goal is to find such $\bar{\alpha}$ that
\begin{equation}
    P[\bar{\alpha}](s) = \frac{\eta_{\rm qsvt}}{\kappa_0}\yFi(s) + \varepsilon_{\rm appr},
\end{equation}
where we set $\kappa_{\rm qsvt} = \kappa_0$, and $\varepsilon_{\rm appr}$ indicates the error in approximating the normalized function $\yFi(s)$ with the polynomial $P[\bar{\alpha}](s)$ [Eq.~\eqref{eq:qsvt-comp}] using the estimated QSVT angles $\bar{\alpha}$.
This expression is similar to that in Eq.~\eqref{eq:P-F-comp}, but here $\varepsilon_{\rm appr} \geq \varepsilon_{\rm qsvt}$ because the estimated angles $\bar{\alpha}$ usually provide worse precision than the original angles $\alpha$.

To compute the metaparameters mentioned above, it is convenient to work with the angles $\theta$.
According to Fig.~\ref{fig:ref-angles}, any two angles $\theta_{j-1}$ and $\theta_j$ have opposite signs (except the case of $j = N_a/2$ where $\theta_{j-1} = \theta_j < 0$). 
We assume that the envelope of the angles $\theta$ does not change with $\kappa_0$ but it is only rescaled by the $\kappa$-dependent maximum amplitude 
\begin{equation}
    \theta_{\rm max}(\kappa) = \max_j|\theta_j(\kappa)|.
\end{equation} 
(By the envelope of QSVT angles, we understand the change of positive and negative normalized angles $\theta_j/\theta_{\rm max}$ with the index $j$.)
Therefore, to estimate $\bar{\theta}$ for an arbitrary $\kappa_0$, one needs to find the functional dependence of $\theta_{\rm max}$ and $N_a$ on the condition number, and describe the envelope of the angles. 
To accomplish this, we use the following technique.
\begin{itemize}
    \item The number $N_a$ for the target $\kappa_0$ is roughly estimated as
        \begin{subequations}\label{eq:Na-est}
        \begin{eqnarray}
            &&N_{a,0} = \floor{N_{a,{\rm ref}}\, \kappa_0 / \kappa_{\rm ref}},\\
            &&N_a = N_{a,0} + {\rm mod}(N_{a,0},2),\label{eq:Na-parity-condition}
        \end{eqnarray}
        \end{subequations}
        where $N_{a,{\rm ref}}$ is the number of the reference angles $\phi_{\rm ref}$ for a single chosen reference value $\kappa_{\rm ref}$.
        Since we use the odd function~\eqref{eq:F} for approximating the inverse function, the estimated $N_a$ should remain even [Eq.~\eqref{eq:Na-Nc}].
        This is ensured by Eq.~\eqref{eq:Na-parity-condition}.
    \item The maximum amplitude of the shifted QSVT angles $\theta$ is approximated in the following way
        \begin{equation}\label{eq:series-ampl}
            \Theta_{\rm max}(\kappa) = \sum_{l = 0}^{N_{\rm ampl} - 1}  \frac{c_{{\rm ampl},l}}{\kappa^l},
        \end{equation} 
    where $N_{\rm ampl}$ is a chosen number of the real scalar metaparameters $c_{{\rm ampl},l}$.
    Although Fig.~\ref{fig:ref-ampl} suggests that the optimal number of $c_{{\rm ampl},l}$ is $N_{\rm ampl} = 2$, numerical tests indicate that it is better to take $N_{\rm ampl} = 5$.
    The scalars $c_{{\rm ampl},l}$ are computed by minimizing the following loss function
        \begin{equation}\label{eq:L-ampl}
            L_{\rm ampl} = \sum_{j=0}^{N_{\rm ref}-1} |\Theta_{\rm max}(\kappa_{{\rm ref},j}) - \theta_{\rm max}(\kappa_{{\rm ref},j})|^2,
        \end{equation}
    where $N_{\rm ref}$ sets of high-precision QSVT angles pre-calculated for $N_{\rm ref}$ values $\kappa_{{\rm ref},j}$ are used.
    \item Then, the angles are normalized by their maximum absolute value:
        \begin{equation}\label{eq:norm-angles}
            \tilde{\theta}_j = \frac{\theta_j}{\theta_{\rm max}},\quad j = 0,1,\dots N_a-1.  
        \end{equation}
    \item The first half of the angles $\tilde{\theta}$ is a mirror reflection of the second half (Fig.~\ref{fig:ref-angles}).
    Therefore, we consider $\tilde{\theta}_j$ only with $j = 0,1,\dots N_a/2-1$ which are split into 
    \begin{subequations}\label{eq:Npos-Nneg}
    \begin{eqnarray}
        &&N_{\rm pos} = \floor{N_a/4},\\
        &&N_{\rm neg} = N_a/2 - \floor{N_a/4}
    \end{eqnarray}
    \end{subequations}
    positive and negative angles, $\tilde{\theta}_j^{\rm pos}$ and $\tilde{\theta}_j^{\rm neg}$, correspondingly.  
    This means that if ${\rm mod}{(N_a, 4)} = 0$, then the number of positive and negative angles is equal.
    \item After that, we attribute a function $G_{\rm pos}(r_j)$ to the normalized positive angles such that
    \begin{eqnarray}\label{eq:G-pos}
        G_{\rm pos}(r_j) = \tilde{\theta}_j^{\rm pos},
    \end{eqnarray}
    where $r_j = \Delta r j$ and $\Delta r = 1/(N_{\rm pos} - 1)$, and assume that the function is even if one extends $r$ from $0$ to $-1$.
    The same procedure is performed for $\tilde{\theta}_j^{\rm neg}$.
    After the mapping ~\eqref{eq:G-pos}, we assume that the angle envelope described by the functions $G_{\rm pos}(r_j)$ and $G_{\rm neg}(r_j)$ does not depend on the condition number.
    \item Then, $G_{\rm pos}(r_j)$ is approximated as
    \begin{equation}\label{eq:series-env-pos}
        \bar{G}_{\rm pos}(r_j) = \sum_{l = 0}^{N_{\rm sh} - 1} c^{\rm pos}_{{\rm sh},l} \cos((2l)\, \arccos(r_j)),
    \end{equation}
    where we assume that the real scalar metaparameters $c^{\rm pos}_{{\rm sh},l}$ do not depend on the condition number, and $N_{\rm sh}$ is a user-defined integer.
    The coefficients $c^{\rm pos}_{{\rm sh},l}$ are computed by minimising the following loss function 
    \begin{eqnarray}\label{eq:L-sh}
        L_{\rm sh} = \sum_{j=0}^{N_{\rm pos}-1} |\bar{G}_{\rm pos}(r_j) - G_{\rm pos}(r_j)|^2,
    \end{eqnarray}
    where $G_{\rm pos}(r_j)$ is computed using $\tilde{\theta}_{\rm ref}$ in Eqs.~\eqref{eq:norm-angles} and~\eqref{eq:G-pos}.
    The negative envelope $\bar{G}_{\rm neg}(r_j)$ is found in the same way.
    The resulting coefficients $c^{\rm neg}_{{\rm sh},l}$ and $c^{\rm pos}_{{\rm sh},l}$ for each $l$ have opposite signs and may differ slightly in their absolute amplitudes. 
    Numerical tests show that $N_{\rm sh} = 20$ is sufficient, and further increasing $N_{\rm sh}$ does not significantly decrease $\varepsilon_{\rm appr}$.
\end{itemize}

\begin{table}
\caption{\label{table:algorithm} 
Summary of the numerical technique for computing the estimated QSVT angles $\bar{\phi}$ for the target value $\kappa_0$.
}
\begin{tabularx}{0.90\textwidth}{|>{\raggedright\arraybackslash}X|}
\hline
    \textit{Computing metaparameters from the reference high-precision QSVT angles.} 
    \begin{enumerate}
    \item[I.1] Compute $N_{\rm ref}$ sets of high-precision QSVT angles for various $\kappa_{{\rm ref},j}$ with $j = 0,1,\dots (N_{\rm ref}-1)$ by using the minimization~\eqref{eq:loss-function}.
    \item[I.2] Find $c_{{\rm ampl},l}$ using Eq.~\eqref{eq:L-ampl}.
    \item[I.3] Choose a single set of high-precision QSVT angles, $\phi_{\rm ref}$, for some $\kappa_{\rm ref}$.
    It is better to take $\phi_{\rm ref}$ for $\kappa_{\rm ref} = \max_j \kappa_{{\rm ref},j}$.
    \item[I.4] Compute $c_{{\rm sh},l}^{\rm pos}$ and $c_{{\rm sh},l}^{\rm neg}$ using Eq.~\eqref{eq:L-sh}.
    \end{enumerate}\\
\hline
    \textit{Estimating QSVT angles using the calculated metaparameters.}
    \begin{enumerate}
    \item[II.1] Choose the target value $\kappa_0$.
    \item[II.2] Compute $N_a$ for the chosen $\kappa_0$ using Eqs.~\eqref{eq:Na-est}.
    \item[II.3] Find the positive and negative envelopes of the estimated angles by using Eq.~\eqref{eq:series-env-pos}.
    Mirror the copies of the envelopes and concatenate them with the original ones to obtain the full envelope.
    \item[II.4] Find the maximum value of the estimated shifted angles using Eq.~\eqref{eq:series-ampl}.
    \item[II.5] Compute the values of the estimated QSVT angles $\bar{\phi}$ by using Eqs.~\eqref{eq:est-theta} and~\eqref{eq:theta}.
    \end{enumerate}\\
\hline
\end{tabularx}
\end{table}

Using $N_{\rm ampl} + 2 N_{\rm sh} + 2$ metaparameters, including the reference value $\kappa_{\rm ref}$, the number $N_{a,{\rm ref}}$ of the reference QSVT angles $\phi_{\rm ref}$, the coefficients $c_{{\rm ampl},l}$, and the coefficients $c^{\rm pos}_{{\rm sh},l}$ and $c^{\rm neg}_{{\rm sh},l}$, one can now estimate the QSVT angles $\bar{\theta}$ for the target value $\kappa_0$.
\begin{itemize}
    \item First of all, one estimates $N_a$ for the chosen $\kappa_0$ using Eq.~\eqref{eq:Na-est} and computes $N_{\rm pos}$ and $N_{\rm neg}$ using Eq.~\eqref{eq:Npos-Nneg}.
    \item After that, the $\kappa$-independent coefficients $c^{\rm pos}_{{\rm sh},l}$ and $c^{\rm neg}_{{\rm sh},l}$ computed in Eq.~\eqref{eq:L-sh} are employed in Eq.~\eqref{eq:series-env-pos} to build the envelope of the estimated positive and negative angles, i.e. $\tilde{\bar{\theta}}^{\rm pos}_j$ with $j = 0, 1,\dots N_{\rm pos} - 1$ and $\tilde{\bar{\theta}}^{\rm neg}_k$ with $k = 0, 1,\dots N_{\rm neg} - 1$, correspondingly.
    \item The order of the elements in the copies of $\tilde{\bar{\theta}}^{\rm pos}$ and $\tilde{\bar{\theta}}^{\rm neg}$ is inverted and the inverted copies are concatenated with the original $\tilde{\bar{\theta}}^{\rm pos}$ and $\tilde{\bar{\theta}}^{\rm neg}$, respectively, to recreate the entire range of the normalized estimated angles $\tilde{\bar{\theta}}_j$ with $j = 0, 1,\dots N_a - 1$.
    \item Using the coefficients $c_{{\rm ampl},l}$ computed in Eq.~\eqref{eq:L-ampl}, one finds the maximum value $\bar{\theta}_{\rm max}$ for the chosen $\kappa_0$ using Eq.~\eqref{eq:series-ampl}.
    \item The final step is to renormalize $\tilde{\bar{\theta}}$ by $\bar{\theta}_{\rm max}$,
    \begin{equation}\label{eq:est-theta}
        \bar{\theta} = \bar{\theta}_{\rm max} \tilde{\bar{\theta}}
    \end{equation}
    which results in the estimated shifted angles $\bar{\theta}$ for the chosen $\kappa_0$.
    The estimated angles $\bar{\phi}$, which are used by the QSVT circuit, can be found from $\bar{\theta}$ using Eq.~\eqref{eq:theta}.
\end{itemize}
The algorithm is summarized in Table~\ref{table:algorithm}.
The code for estimating QSVT angles can be found in~\cite{code-estimation}.

\section{Numerical tests}\label{sec:tests}
To verify the technique described in Sec.~\ref{sec:approach} and compute the error $\varepsilon_{\rm appr}$, we perform several numerical tests using the estimated QSVT angles $\bar{\phi}$. 
These tests include inverting the approximate inverse function~\eqref{eq:F} and a sine function, and simulating a boundary-value problem that describes the linear dynamics of EM waves in a dielectric medium.
In all these tests, we use $\kappa_{\rm ref} = 650$ to compute the envelope coefficients $c^{\rm pos}_{{\rm sh},l}$ and $c^{\rm neg}_{{\rm sh},l}$ in Eq.~\eqref{eq:L-sh}.
Also, we use $N_{\rm ref} = 65$ sets of high-precision angles with $\kappa_{\rm ref} = 10, 20,\dots 650$ to compute the coefficients $c_{{\rm ampl},l}$ in Eq.~\eqref{eq:L-ampl}.
The simulated test cases and the corresponding quantum circuits can be found in~\cite{code-estimation}.

\subsection{Inverting $\yFi$ and sine functions}
To verify that the QSVT angles are estimated correctly, we invert the normalized function~\eqref{eq:F} for various $\kappa_{\rm qsvt}$. 
More precisely, we block encode the following diagonal matrix 
\begin{equation}\label{eq:A-inv-F}
    A_{\yFi} = \frac{\eta_{\rm A}}{\kappa_{\rm qsvt}} {\rm diag}(\yFi_0, \yFi_1, ... \yFi_{N_x-1})
\end{equation}
where $\yFi_k = \yFi(x_k)$ with $x_k$ is a grid with $N_x = 2^{n_x}$ points satisfying Eq.~\eqref{eq:interv}, and $\eta_{\rm A} \leq 1$ is a rescaling factor of the spectral norm of the matrix $A_{\yFi}$.
The matrix $A_{\yFi}$ has singular values 
\begin{equation}\label{eq:s-A-inv}
    s_k = \left|\frac{\eta_{\rm A} \yFi_k}{\kappa_{\rm qsvt}}\right|,
\end{equation}
and its condition number $\rho_A$ is equal to $\kappa_{\rm qsvt}$ if $\eta_A = 1$.
The maximum singular value is equal to $\eta_A$.
The normalization to $\kappa_{\rm qsvt}$ is necessary to make the matrix spectral norm less or equal to one which is a necessary condition for the block-encoding of $A_{\yFi}$.
Then, we invert $A_{\yFi}$ by using circuit~\ref{circ:qsvt}.
There, we use the estimated QSVT angles $\bar{\phi}$ computed with the algorithm~\ref{sec:approach} by setting $\kappa_0 = \rho_A$.

\begin{figure}[!t]
\centering
\subfloat{\includegraphics[width=0.49\textwidth]{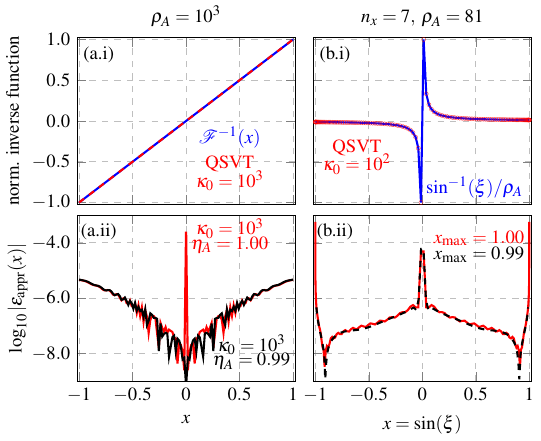}}
\subfloat{\includegraphics[width=0.49\textwidth]{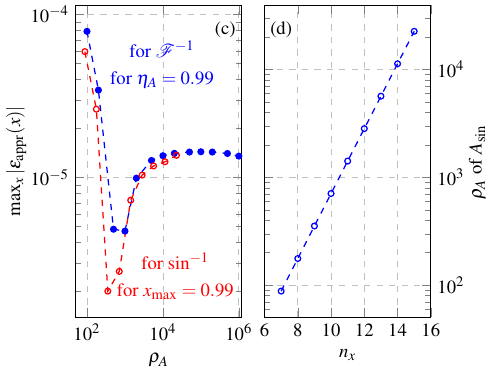}}
\caption{
    \label{fig:Dir-x} 
    (a): Plots showing comparison of the inverse function $\yFi^{-1}(x)$ computed classically (blue line) and by QSVT (red line) for $\rho_A = 10^3$ (a.i) and the corresponding QSVT error (a.ii) with $\eta_A = 1.00$ (red line) and $\eta_A = 0.99$ (black line). 
    (b): Comparison of the inverse function $\sin^{-1}(\xi)$ computed classically (blue line) and by the QSVT (red markers) for $n_x = 7$ and $\rho_A = 81$ (b.i) and the corresponding QSVT error (b.ii) with $x_{\rm max} = 1.00$ (red line) and $x_{\rm max} = 0.99$ (black line).
    (c): The maximum QSVT error for various $\rho_A$ in the computation of the inverse functions $\yFi^{-1}(x)$ (blue markers) and $\sin^{-1}(\xi)$ (red markers).
    (d): The dependence of the condition number $\rho_A$ of the matrix $A_{\sin{}}$ on $n_x$.
}
\end{figure}

According to Eq.~\eqref{eq:qsvt-output-res}, the result of the inversion of Eq.~\eqref{eq:A-inv-F} by QSVT is a state vector with the elements 
\begin{equation}
    \psi_{{\rm in}, k} = \frac{\eta_{\rm qsvt} x_k}{\eta_A 2^{n_x/2}},
\end{equation}
where the term $2^{-n_x/2}$ appears from $\ket{{\rm init}}_{\rm in}$, and we do not take into account the global phase.
After renormalizing $\ket{\psi}_{\rm in}$ by $2^{n_x/2}/\eta_{\rm qsvt}$, this vector encodes a linear function from $-\eta_A^{-1}$ to $\eta_A^{-1}$. 
The comparison between the classical and QSVT results is shown in Fig.~\ref{fig:Dir-x}a.
The numerical tests show that the maximum error is localized at the singular value $s = 1$.
This error can be easily avoided by setting $\eta_A$ to a value less than one.
In fact, it is often difficult to block-encode a matrix while keeping its spectral norm equal to one.
Therefore, this localized error should not be noticeable in most QSVT simulations.
According to Fig.~\ref{fig:Dir-x}c, the maximum absolute error of the QSVT computation of the normalized inverse function is around $10^{-5}$ for large condition numbers.
The drop of the error at $\rho_A \approx 10^3$ is observed because $\rho_A$ becomes close to the $\kappa_{\rm ref} = 650$ used for computing the envelope coefficients of the reference QSVT angles $\phi_{\rm ref}$.

As another test, we invert the matrix 
\begin{subequations}\label{eq:test-sin}
\begin{eqnarray}
    &&A_{\sin{}} = {\rm diag}(\sin\xi_k),\\
    &&\xi_k = -\xi_{\rm max} + \frac{2 \xi_{\rm max} k}{N_x - 1},\quad k = 0, 1,... (N_x-1),
\end{eqnarray}
\end{subequations}
whose condition number $\rho_A$ can be increased by increasing $n_x$ as one can see from Fig.~\ref{fig:Dir-x}d. 
The singular values of $A_{\sin{}}$ are
\begin{equation}\label{eq:s-A-sin}
    s_k = |\sin\xi_k|.
\end{equation}
If $\xi_{\rm max} = \pi/2$, the spectral norm of $A_{\sin{}}$ is one.
To compute $A_{\sin{}}^{-1}$ using the QSVT circuit, we estimate the QSVT angles $\bar{\phi}$ by setting $\kappa_0 \geq \rho_A$.
Similarly to the previous case, the maximum error of the QSVT computations with $\bar{\phi}$ is localized at $s = 1$ (Fig.~\ref{fig:Dir-x}b).
By slightly decreasing $\xi_{\rm max}$, one can set $||A_{\sin{}}|| < 1$ and, due to this, reduce the maximum error to $10^{-5}$.
The dependence of $\varepsilon_{\rm appr}$ on $\rho_A$ for both $\yFi^{-1}(x)$ and $\sin^{-1}(\xi)$ shows the same behaviour as one can see from Fig.~\ref{fig:Dir-x}c.

\subsection{Electromagnetic (EM) waves}\label{sec:EM}
\begin{figure}[!t]
\centering
\subfloat{\includegraphics[width=0.49\textwidth]{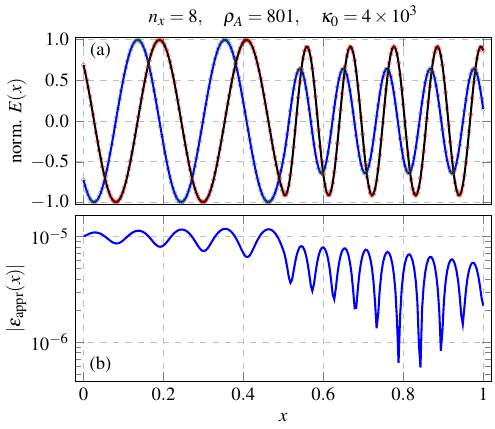}}
\subfloat{\includegraphics[width=0.49\textwidth]{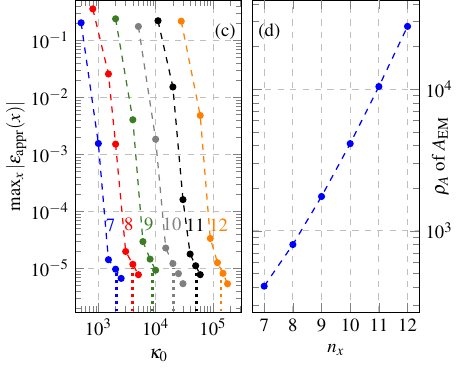}}
\caption{
    \label{fig:EM-x} 
    (a): Plot showing the real (solid blue line) and imaginary (solid black line) components of the electric field $E$ computed classically, and the real (green markers) and imaginary (red markers) components of $E$ computed by the QSVT circuit~\ref{circ:qsvt} with $\bar{\phi}$ estimated for $\kappa_0 = 4\times 10^3$. Here, $n_x = 8$ and $\rho_A = 801$.
    (b): The difference between the classical and QSVT signals.
    (c): The dependence of $\max_x|\varepsilon_{\rm appr}(x)|$ on the value of $\kappa_0$ used for the estimation of $\bar{\phi}$ for various $n_x$, which are indicated by colored integers.
    The colored vertical dashed lines indicate the values $5 \rho_A$ for various $n_x$.
    (d): The dependence of the condition number $\rho_A$ of the matrix $A_{\rm EM}$ on $n_x$.
}
\end{figure}

Here, we consider a one-dimensional boundary-value problem describing stationary EM waves,
\begin{subequations}\label{sys:model}
\begin{eqnarray}
    &&\yi\omega\epsilon_L E(x) + \partial_x B(x) = 0,\label{eq:model-E}\\
    &&\yi\omega B(x) + \partial_x E(x) = 0,\label{eq:model-B}
\end{eqnarray}
\end{subequations}
with the outgoing boundary conditions at $x = 0$ and $x = 1$ and the source $Q = Q_0 \exp(\yi\omega t)$:
\begin{subequations}\label{eq:model-boundaries}
\begin{eqnarray}
    &&(\yi\omega - \partial_x) E\big|_{x = 0} = 0,\label{eq:left-cond}\\
    &&(\yi\omega + \partial_x) B\big|_{x = 1} = Q_0,\label{eq:right-cond}
\end{eqnarray}
\end{subequations}
on a spatial grid $x = [0,1]$ with $N_x = 2^{n_x}$ points in a medium with the following dielectric permittivity
\begin{equation}\label{eq:epsilon-rint}
\epsilon_L
    = \left\{
    \begin{array}{cc}
        \epsilon_0, & x < 0.5,\\
        \epsilon_1, & x > 0.5.
    \end{array}
    \right.
\end{equation}
The detailed description of the system and the corresponding matrix $A_{\rm EM}$ encoding the EM problem are described in~\cite{Novikau23}.
We use this problem as another test case for the verification of the estimated QSVT angles $\bar{\phi}$ for various $\kappa_0$.
The comparison between classical and QSVT simulations of this dynamical system in the case with $n_x = 8$ is shown in Fig.~\ref{fig:EM-x} where one can see the spatial distribution of the real and imaginary components of the stationary electric field $E$. 
The absolute difference between the classical and QSVT signals is around $10^{-5}$.
As shown in Fig.~\ref{fig:EM-x}d, the condition number $\rho_A$ of the matrix $A_{\rm EM}$ encoding the EM wave problem grows with increasing $n_x$, and for $n_x = 8$, it is equal to $801$.
The spectral norm $||A_{\rm EM}||$ of the block-encoded matrix is $0.2$ and practically does not depend on $n_x$.
Hence, according to Eq.~\eqref{eq:kappa-qsvt-norm} and Fig.~\ref{fig:EM-x}c, one should take $\kappa_0 \approx ||A_{\rm EM}||^{-1} \rho_A$ to minimize the QSVT error.
With this condition satisfied, the maximum error of the QSVT computation with the estimated angles $\bar{\phi}$ is around $10^{-5}$ for all simulated $n_x$.

\subsection{Discussion}\label{sec:discussion}

\begin{figure}[!t]
\centering
\includegraphics[width=0.98\textwidth]{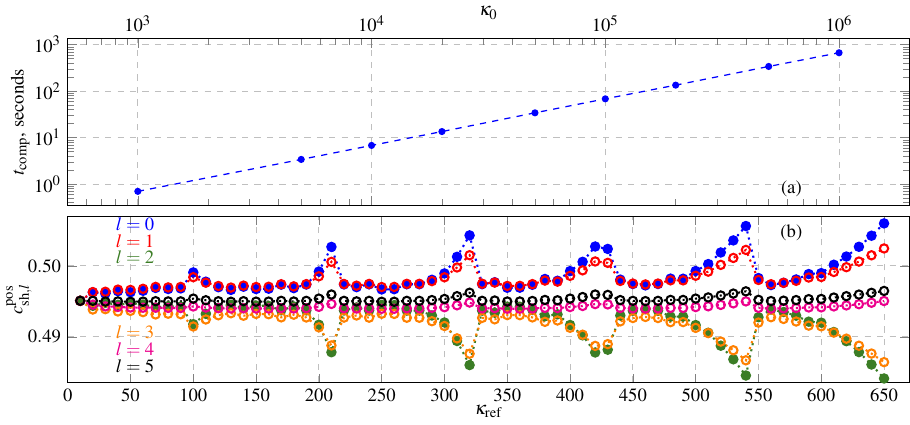}
\caption{
    \label{fig:Scaling_time_kappa} 
    (a) The dependence of the computational time $t_{\rm comp}$ used by the approach~\ref{sec:approach} for estimating $\bar{\phi}$ for various $\kappa_0$.
    (b) The dependence of the coefficients $c^{\rm pos}_{{\rm sh},l}$ [Eq.~\eqref{eq:series-env-pos}] for several $l$ on $\kappa_{\rm ref}$.
    Here, for the sake of clearness, all $c^{\rm pos}_{{\rm sh},l}$ with $l \neq 1$ are shifted towards $c^{\rm pos}_{{\rm sh},1}$ in the following way: $c^{\rm pos}_{{\rm sh},l}(\kappa_{\rm ref}) = c^{\rm pos}_{{\rm sh},l}(\kappa_{\rm ref}) + \left[c^{\rm pos}_{{\rm sh},1}(10) - c^{\rm pos}_{{\rm sh},l}(10)\right]$.  
}
\end{figure}

According to the numerical tests discussed in the previous sections, one can expect $\varepsilon_{\rm appr} \approx 10^{-5}$ in the QSVT computations with the estimated angles $\bar{\phi}$ with $\kappa_0 \leq 10^{6}$.
The computational time required for estimating $\bar{\phi}$ grows linearly with $\kappa_0$ as shown in Fig.~\ref{fig:Scaling_time_kappa}a.
To improve the precision of the proposed estimation algorithm, one can optimize $N_a$ using the approximation~\eqref{eq:Na-est} as an initial guess.
Indeed, according to Fig.~\ref{fig:Dir-x}c, the error $\varepsilon_{\rm appr}$ increases when one computes $\bar{\phi}$ for small $\kappa_0$.
This may mean that there could be a constant factor in the scaling~\eqref{eq:Na-est}.
The contribution of this term to $N_a$ is stronger for small $\kappa_0$.
We are not interested in estimating $\bar{\phi}$ for $\kappa_0 \leq 650$ because one can relatively easy compute $\phi$ for low condition numbers using the high-precision algorithms~\cite{Dong21, Ying22}.
Yet, by optimizing the estimation of $N_a$ for low $\kappa_0$, one can find the missing constant factor in Eq.~\eqref{eq:Na-est} and, thus, can potentially decrease  $\varepsilon_{\rm appr}$ for large $\kappa_0$ as well.
Moreover, according to Fig.~\ref{fig:ref-Na}, $N_a$ does not grow exactly linearly with the condition number.
This results in an additional error when one uses the linear scaling~\eqref{eq:Na-est}.

Another option to improve the precision of the proposed estimation technique is to take into account the dependence of the Chebyschev coefficients $c^{\rm pos}_{{\rm sh},l}$ and $c^{\rm neg}_{{\rm sh},l}$ on the condition number.
In the current version of the algorithm, we compute $c_{{\rm sh},l}$ for a single $\kappa_{\rm ref} = 650$.
However, $c_{{\rm sh},l}$ for each particular $l = 0, 1,\dots (N_{\rm sh}-1)$ slightly depends on the condition number as one can see in Fig.~\ref{fig:Scaling_time_kappa}b, which shows the dependence of a few first coefficients $c^{\rm pos}_{{\rm sh},l}$ on $\kappa_{\rm ref}$.
One can reformulate the minimization problem~\eqref{eq:loss-function} in terms of $2N_{\rm sh}$ coefficients $c_{{\rm sh},l}$ instead of $N_a$ angles $\alpha_k$.
As a result, instead of optimizing $N_a$ QSVT angles for the chosen $\kappa_0$ (e.g., $N_a \approx 4 \times 10^7$ for $\kappa_0 = 10^6$), one will need to minimize a loss function with respect to $2N_{\rm sh} \approx 40$ real parameters $c_{{\rm sh},l}$.
As an option, the FFPI technique recently developed in Ref.~\cite{Ni24} can be used to efficiently generate more data for training the parameters $c_{{\rm sh},l}$ to take into account their dependence on the condition number.
That said, such optimization will be more expensive than the current fast estimation technique, and it may be preferable to use the FFPI algorithm~\cite{Ni24} to compute high-precision QSVT angles directly, although the latter has not yet been tested for such large condition numbers considered in this work.

\section{Conclusions}\label{sec:conclusion}
We have proposed a new numerical technique for estimating QSVT angles for matrix inversion with large condition numbers. 
The algorithm is based on using a small number of metaparameters that describe the dependence of QSVT angles on the condition number.
By operating with these metaparameters instead of the angles directly, we have significantly reduced the time necessary to compute the QSVT angles while maintaining relatively high precision. 
The computational time of this algorithm scales linearly with the target condition number. 
The estimated angles can be used in QSVT circuits to solve systems of linear equations with an error of $10^{-5}$ for condition numbers as large as $10^6$.
The precision of the technique can potentially be increased by optimizing the metaparameter values for the target condition numbers.

\section*{Acknowledgments}
The work, LLNL-JRNL-868744, was supported by the U.S. Department of Energy (DOE) Office of Fusion Energy Sciences “Quantum Leap for Fusion Energy Sciences” Project No. FWP-SCW1680 at Lawrence Livermore National Laboratory (LLNL). 
Work was performed under the auspices of the U.S. DOE under LLNL Contract DE-AC52–07NA27344.
This research used resources of the National Energy Research Scientific Computing Center, a DOE Office of Science User Facility supported by the Office of Science of the U.S. Department of Energy under Contract No. DE-AC02-05CH11231 using NERSC award FES-ERCAP0028618.

\bibliography{main}
\end{document}